%
\let\useblackboard=\iftrue
%
\let\useblackboard=\iffalse
%
%
\input harvmac.tex
%
\input epsf.tex
\ifx\epsfbox\UnDeFiNeD\message{(NO epsf.tex, FIGURES WILL BE
IGNORED)}
\def\figin#1{\vskip2in}
\else\message{(FIGURES WILL BE INCLUDED)}\def\figin#1{#1}\fi
\def\ifig#1#2#3{\xdef#1{fig.~\the\figno}
\midinsert{\centerline{\figin{#3}}%
\smallskip\centerline{\vbox{\baselineskip12pt
\advance\hsize by -1truein\noindent{\bf Fig.~\the\figno:} #2}}
\bigskip}\endinsert\global\advance\figno by1}
\noblackbox
\baselineskip=12pt
\useblackboard
\message{If you do not have msbm (blackboard bold) fonts,}
\message{change the option at the top of the tex file.}

\font\blackboard=msbm10 scaled \magstep1
\font\blackboards=msbm7
\font\blackboardss=msbm5
\textfont\black=\blackboard
\scriptfont\black=\blackboards
\scriptscriptfont\black=\blackboardss

\else

\fi

%
\def\yboxit#1#2{\vbox{\hrule height #1 \hbox{\vrule width #1
\vbox{#2}\vrule width #1 }\hrule height #1 }}
\def\fillbox#1{\hbox to #1{\vbox to #1{\vfil}\hfil}}
\def\ybox{{\lower 1.3pt \yboxit{0.4pt}{\fillbox{8pt}}\hskip-0.2pt}}

\def\comments#1{}

\def\CN{{\cal N}}

\def\II{\relax{I\kern-.07em I}}

\def\IZ{\relax\ifmmode\mathchoice
{\hbox{\cmss Z\kern-.4em Z}}{\hbox{\cmss Z\kern-.4em Z}}
{\lower.9pt\hbox{\cmsss Z\kern-.4em Z}}
{\lower1.2pt\hbox{\cmsss Z\kern-.4em Z}}\else{\cmss Z\kern-.4em
Z}\fi}

\def\nfour{$\CN \!=\! 4$}
\def\neight{$\CN \!=\! 8$}

%
%

%
%
\lref\juani{J. Maldacena, ``The Large N Limit of Superconformal Field Theories
and Supergravity'', Adv. Theor. Math. Phys. 2 (1998) 231, hep-th/9711200.}
\lref\gkp{ S. S. Gubser, I. R. Klebanov and A. M. Polyakov, ``Gauge Theory 
Correlators from Non-Critical String Theory'', Phys. Lett. B428 (1998) 105, 
hep-th/9802109.}

\lref\witi{E. Witten, ``Anti De Sitter Space And Holography'', 
Adv. Theor. Math. Phys. 2 (1998) 253, hep-th/9802150.}
\lref\witii{E. Witten, ``Anti-de Sitter Space, Thermal Phase Transition, 
And Confinement In Gauge Theories'', Adv. Theor. Math. Phys. 2 (1998) 505,
hep-th/9803131.}
\lref\hp{S. W. Hawking and D. Page, ``Thermodynamics of Black Holes
in Anti-de Sitter Space'', Comm. Math. Phys. 87 (1983) 577.}
\lref\rcm{R. Emparan, C. V. Johnson and R. C. Myers, ``Surface Terms as 
Counterterms in the AdS/CFT Correspondence'', hep-th/9903238.}
\lref\review{O. Aharony, S. S. Gubser, J. Maldacena, H. Ooguri and Y. Oz,
``Large N Field Theories, String Theory and Gravity'', hep-th/9905111.}
\lref\bcm{C. P. Burgess, N. R. Constable and R. C. Myers, ``The Free Energy of N=4 Super-Yang-Mills
Theory and the AdS/CFT Correspondence'', hep-th/9907188.}
\lref\stva{ A. Strominger and C. Vafa, ``Microscopic Origin of the Bekenstein-Hawking Entropy'', Phys. Lett. B379 (1996) 99, hep-th/9601029.}
\lref\garyi{ G. T. Horowitz and S. F. Ross, ``Possible Resolution of Black Hole Singularities from Large N Gauge Theory'', JHEP 9807 (1998) 014,
hep-th/9803085.}
\lref\toni{I. T. Drummond, R. R. Horgan, P. V. Landshoff and A. Rebhan, ``Foam 
Diagram Summation at Finite Temperature'', Nucl. Phys. B524 (1998) 579, hep-ph/9708426.}
\lref\itzhaki{ N. Itzhaki, ``A Comment on the Entropy of Strongly Coupled N=4'',
hep-th/9904035.}
\lref\gkt{S. S. Gubser, I. R. Klebanov and A. A. Tseytlin, ``Coupling Constant Dependence in the Thermodynamics of N=4 Supersymmetric Yang-Mills Theory'', Nucl. Phys. B534 (1998) 202, hep-th/9805156.}
\lref\weak{A. Fotopoulos and T. R. Taylor, `` Remarks on Two-Loop Free Energy 
in N=4 Supersymmetric Yang-Mills Theory at Finite Temperature'', Phys. Rev. D59 (1999) 61701, hep-th/9811224;
 C. Kim and S.-J. Rey, ``Thermodynamics of Large-N Super Yang-Mills Theory and AdS/CFT Correspondence'', hep-th/9905205;
M. A. Vazquez-Mozo, ``A Note on Supersymmetric Yang-Mills Thermodynamics'',
hep-th/9905030;
A. Nieto and M. H. G. Tytgat, ``Effective Field Theory Approach to N=4 Supersymmetric Yang-Mills at Finite Temperature'', hep-th/9906147.}
\lref\old{M. B. Altaie and J. S. Dowker, ``Spinor Fields in an Einstein
Universe: Finite-temperature effects'', Phys. Rev. D18 (1978) 3557.}
\lref\myers{C. P. Burgess, N. R. Constable and R. C. Myers, ``The Free Energy of N=4 Super-Yang-Mills and the AdS/CFT Correspondence'', JHEP 9908 (1999) 017,
hep-th/9908175.}
\lref\hawkingi{S. W. Hawking, C. J. Hunter and M. M. Taylor-Robinson, ``Rotation and the AdS/CFT Correspondence'', Phys. Rev. D59 (1999) 064005, hep-th/9811056.}
\lref\hawkingii{S. W. Hawking and H. S. Reall, ``Charged and Rotating AdS Black Holes and Their CFT Duals'', hep-th/9908109.}
\lref\berpar{D. S. Berman and M. K. Parikh, ``Holography and Rotating AdS 
Black Holes'', hep-th/9907003.}
\lref\bala{V. Balasubramanian and P. Kraus,''A Stress Tensor for Anti-de Sitter Gravity'', hep-th/9902121.}
\lref\cliff{A. M. Awad and C. V. Johnson, `` Holographic Stress Tensors for Kerr-AdS Black Holes'', hep-th/9910040.}
\lref\carter{B. Carter, ``Hamilton-Jacobi and Schrodinger Separable Solutions 
of Einstein's Equations'', Commun. Math. Phys. 10, (1968) 280.}  
\lref\kt{I. R. Klebanov and A. A. Tseytlin, ``Entropy of Near-Extremal Black p-branes'', Nucl.Phys. B475 (1996) 179, hep-th/9604166.}
\lref\gka{S. S. Gubser, I. R. Klebanov and A. W. Peet, ``Entropy and 
Temperature of Black 3-Branes'', Phys.Rev. D54 (1996) 3915, hep-th/9602135.}
\lref\kl{Y. Gao and M. Li, `` Large N Strong/Weak Coupling Phase Transition and the Correspondence Principle'', Nucl.Phys. B551 (1999) 229, hep-th/9810053;
K. Landsteiner, ``String Corrections to the Hawking-Page Phase Transition'', Mod. Phys. Lett. A14 (1999) 379, hep-th/9901143;
M. M. Caldarelli and D. Klemm, ``M-Theory and Stringy Corrections to Anti-de Sitter Black Holes and Conformal Field Theories'', Nucl. Phys. B555 (1999) 157,
hep-th/9903078.}
\lref\ckc{M. M. Caldarelli, G. Cognola and D. Klemm, ``Thermodynamics of Kerr-Newman-AdS Black Holes and Conformal Field Theories'', hep-th/9908022.}
\lref\bo{B. Sundborg, ``The Hagedorn Transition, Deconfinement and N=4 SYM Theory'' ,hep-th/9908001.}
\lref\agmoo{O. Aharony, S. S. Gubser, J. Maldacena, H. Ooguri and Y. Oz,
``Large N Field Theories, String Theory and Gravity'', hep-th/9905111.}
\lref\hoso{Y. Hosotani, ``Dynamical mass generation in SU(N)
Coleman-Weinberg
gaue theory in the Einstein static universe'', Phys. Rev. D30 (1984)
1238.}
%
\Title{ \vbox{\baselineskip12pt\hbox{hep-th/9911124}
\hbox{CERN-TH-99-345}
}}
{\vbox{
\centerline{The Thermodynamic Potentials of}
\centerline{Kerr-AdS Black Holes and their CFT Duals} }}

\centerline{Karl Landsteiner and Esperanza Lopez}
\medskip
\centerline{Theory Division CERN}
\centerline{CH-1211 Geneva 23}
\centerline{Switzerland}
\medskip
\centerline{\tt Karl.Landsteiner@cern.ch}
\centerline{\tt Esperanza.Lopez@cern.ch}

\vskip15mm

\centerline{\bf Abstract}
String or M-theory in the background of Kerr-AdS black holes is thought 
to be dual to the large $N$ limit of certain
conformal field theories on a rotating sphere at finite temperature.
The five dimensional black hole is associated to 
\nfour\ supersymmetric Yang-Mills theory on a rotating three-sphere and 
the four dimensional one to the superconformal field theory of 
coinciding M2 branes
on a rotating two-sphere.
The thermodynamic potentials can be expanded in 
inverse powers of the radius of the sphere.
We compute the leading and subleading terms of this expansion
in the field theory at one loop and compare
them to the corresponding supergravity expressions.
The ratios between these terms at weak and strong coupling turns
out not to depend on the rotation parameters in the case of 
\nfour\ SYM.
For the field theory living on one M2 brane
we find a subleading logarithmic term.
No such term arises from the supergravity calculation.
\baselineskip=14pt

\vfill
\Date{\vbox{\hbox{\sl November 1999}}}

\newsec{Introduction}

The AdS/CFT correspondence relates string or
M-theory on $M_n \times X^{D-n}$, where $M_n$ is an space of negative curvature and $X^{D-n}$ an Einstein manifold, to field theories living
on the (conformal) boundary of these spaces \refs{\juani,\gkp,\witi} (see
\agmoo\ for a comprehensive review). 
It is particularly interesting to consider $M_n$ being asymptotically AdS
black holes.
The dual field theories are then at finite temperature, with
the field theory temperature given by the Hawking temperature of the
black hole.

An example is the AdS black hole with planar horizon.
It arises as the near horizon
limit of the near extremal D3-brane \garyi. 
The dual field theory is \nfour\ supersymmetric Yang-Mills theory.
It turns out that the entropy, as calculated from the supergravity 
approximation to string theory, differs by a 
factor of $3\over 4$ from the one calculated in the gauge theory \gka.
This difference comes from the fact that the
supergravity side corresponds to the gauge theory at infinite 't Hooft 
coupling, $\lambda=g_{YM}^2 N$, whereas the gauge theory calculation 
applies for weak 't Hooft
coupling\foot{In \itzhaki\ it was argued on general grounds that there should 
be a factor of order one between the strong and weak coupling result in 
\nfour\ Yang-Mills.}$^,$\foot{It is interesting to note that a factor of 
$4\over 5$ 
between the thermal pressure at strong and weak coupling appears in the 
large $N$ limit of an 
$O(N)$ scalar field theory in three dimensions \toni.}.  
The coupling constant dependence of the 
entropy has been investigated in the strong \gkt\ and
weak \weak\ coupling limits.  Because of the underlying conformal
symmetry the temperature is the only scale in these problems. For this reason
it is guaranteed that the entropy of the AdS black hole  scales with the
temperature in the same way as the conformal field theory. It seems therefore
interesting to investigate situations where there are dimensionless parameters.

This can be achieved by considering the conformal field theory
living on a sphere. The entropy can then be multiplied
by an a priori arbitrary function of the product of the temperature and
the radius of the sphere. The supergravity 
duals are
AdS black holes with spherical horizons.  The precise dependence on the
dimensionless parameter $\epsilon ={1\over T R}$ is easy to calculate in 
gravity.
On the conformal field theory side it is however far from trivial to 
extract this dependence in a closed form. Expressions for the
one loop energy of conformal fields on the three-sphere at finite temperature 
have
been obtained long ago in \old\ in form of infinite sums. One can however 
reside to a high temperature or large radius expansion 
\eqn\expansion{F = - V N^2 T^4 \sum_{n=0}^{\infty} b_n(\lambda) \epsilon^n\; .}
The leading term in such an expansion coincides with the flat space result.
A strong weak/coupling comparison of the subleading terms has been 
performed in \myers. The coupling constant dependence on the 
supergravity limit has been investigated in \kl.  

A way to include more dimensionless parameters is to consider Kerr-AdS black 
holes \refs{\hawkingi, \ckc, \hawkingii, \cliff}. The angular velocities give 
rise to
new dimensionless parameters. 
The thermodynamic potential of these black holes presents 
a characteristic divergence when the angular velocity reaches the speed 
of light. 
The AdS correspondence relates these black holes to conformal
field theories on a rotating Einstein universe. 
The thermodynamic potential of the conformal field theory  
shows
the same divergence as its supergravity dual. In \berpar\ it was shown 
that the ratio of
the potentials in the extreme high temperature limit is 
independent
of the angular velocities and still $3\over 4$. A numerical calculation
of the ratio between the free energies of the AdS black hole and the
conformal field theories with varying rotation parameter has been performed 
in \hawkingii. They found that the ratio depends on the rotation
parameters although it always tends to $3\over 4$ in the high temperature 
limit.

The aim of this paper is to calculate the subleading terms in the
high temperature expansion of the thermodynamic potential in field theory with 
rotation parameters present. For the case of \nfour\ SYM on $S^3$ 
we have two independent angular velocities corresponding to the
Cartan subalgebra of the $SU(2)\times SU(2)$ isometries of $S^3$. 
We take the generic case with both angular velocities different from
zero and also different from each other. We find that the functional
dependence of the subleading term is the same at weak and at strong 
coupling. Their ratio is given by ${b_2(\infty)\over b_2(0)} = {3\over 2}$,
which coincides with the result in the non-rotating case \myers.
This is remarkable since there does not seem to exist an obvious symmetry 
that constrains the functional form of this coefficient. Our result is also 
consistent with \hawkingii\ since the
ratio of the thermodynamic potentials at strong and weak coupling does
depend on the angular velocities.

Four dimensional Kerr-AdS black holes are dual to the superconformal 
field theory arising on the world volume of M2 branes on a rotating 
two-sphere. 
We do not have an explicit formulation
of this theory when more that one M2-brane coincide. We know however
the theory for a single M2-brane, it is the 
\neight\ supersingleton theory. M-theory on $AdS_4\times S^7$ has
only one expansion parameter, the number of coincident M2-branes. 
It seems therefore the
most natural analogue to the strong/weak coupling comparisons in four dimensions
to compare the theories in three dimensions at $N=\infty$ and at $N=1$.
We find that the subleading term at $N=1$ behaves logarithmically with
the temperature. This is in strong contrast to the behaviour at $N=\infty$
where we find only polynomial behaviour in the high temperature expansion.

In order to derive subleading contributions to the thermodynamic
potential of the field theory living on spheres, we have to evaluate
sums over modes.
We present a simple method to approximate sums to arbitrary 
accuracy when an small parameter is present. In our case the small parameter 
is ${1\over T R}$. Our method might prove useful in other examples.

\newsec{Supergravity calculation}

\subsec{Summary of thermodynamics of five dimensional Kerr-AdS black holes}
The five dimensional Kerr-AdS metric has been derived and studied in \hawkingi.
We quote briefly some of their results. The metric can be written as
\eqn\kerradsv{\eqalign{
ds^2 =& - {\Delta \over \rho^2} \left( dt - {a_1 \sin^2{\theta} \over \Xi_1} d\phi_1 -
{a_2 \cos^2 \theta \over \Xi_2} d\phi_2 \right)^2 + {\Delta_\theta \sin^2\theta \over
\rho^2} \left( a_1 dt - {(r^2+a_1^2)\over \Xi_1} d\phi_1 \right)^2 + \cr
&+ {\Delta_\theta \cos^2 \theta \over \rho^2} \left( a_2 dt -{(r^2+a_2^2) \over\Xi_2} d\phi_2\right)^2 + {\rho^2\over\Delta} dr^2 +{\rho^2\over\Delta_\theta} d\theta^2 + \cr
&+ {(1+r^2)\over r^2 \rho^2} \left( a_1 a_2 dt - {a_2(r^2+a_1^2) \sin^2\theta\over\Xi_1} d\phi_1 - {a_1(r^2+a_2^2) \cos^2\theta\over\Xi_2} d\phi_2 \right)^2 \,,
}}
where
\eqn\defsv{\eqalign{
\Delta = & \; {1\over r^2} (r^2+a_1^2)(r^2+a_2^2)(1+r^2)-2m\, ,\cr
\Delta_\theta = & \; 1-a_1^2 \cos^2\theta -  a_2^2 \sin^2\theta \, ,\cr
\rho^2 = & \; r^2 + a_1^2 \cos^2 \theta + a_2^2 \sin^2 \theta \, ,\cr
\Xi_i = & \; 1-a_i^2 \, .
}}
The parameter $m$ is related to the black hole mass and
$a_i$ to the angular velocities. The particular case $m=0$
corresponds to empty AdS. The asymptotic AdS nature of the metric
\kerradsv\ can be exhibited by introducing new coordinates
\eqn\newcoorv{\eqalign{
t =& t\, ,\cr
\Xi_1 y^2 \sin^2\Theta = & (r^2+a_1^2)\sin^2\theta\, ,\cr
\Xi_2 y^2 \cos^2\Theta = & (r^2+a_2^2)\cos^2\theta\, ,\cr
\Phi_i = \phi_i+a_i t\, .
}}

The horizon radius $r_+$ is defined as the largest root of $\Delta=0$. 
In the coordinates \kerradsv\ both the horizon and the sphere at infinity
rotate. The angular velocities
\eqn\velosv{ 
\Omega_i = \Omega_i^H - \Omega_i^
{\infty} = {a_i (1+r_+^2)\over r_+^2+a_i^2} \, ,
} 
act as chemical potentials for the angular momenta of the fields
in the dual field theory. 

We have set the scale of AdS to one such that the period of the Euclidean 
time coordinate is 
\eqn\betav{ \beta = {1\over T} = {4\pi (r_+^2+a_1^2)(r_+^2+a_2^2)\over
r_+^2 \Delta^\prime(r_+)\,.}}
With these conventions $\beta={1\over T}$ and $\Omega_i$ are taken to be
dimensionless. The Euclidean action calculated with respect to AdS is
\eqn\actionv{ 
I = -\log Z = - {\pi \beta (r_+^2 + a_1^2)(r_+^2 + a_2^2)(r_+^2-1)
\over 8 G_5 (1-a_1^2)(1-a_2^2)}\,.
}
$G_5$ is the Newton's constant in five dimensions. The thermodynamic potential is
defined by $ F = T I$. Similar as in the non-rotating case
there is a phase transition \hp\ at $r_+=1$.  The field theory 
interpretation is that there is a deconfining phase for $r_+ > 1$ \witii. 
This is also the regime where we want to compare with the field theory at 
weak coupling. 
In order to do so we would like to express the thermodynamic potential in
terms of $T$ and $\Omega_i$. In the large $T$ and large $r_+$ regime we
can invert \velosv\ and \betav\ approximately
\eqn\inversionv{\eqalign{
r_+ = \; & \pi T \, - \, {1-\Omega_1^2-\Omega_2^2\over 2 \pi T} \, + 
O\biggl( { 1\over T^2}\biggr)\, ,\cr
a_i = \; & \Omega_i \left( 1 - {1-\Omega_i^2\over\pi^2 T^2} \right) \,
+ \, O\biggl( {1\over T^3} \biggr) \, .}
}
Using these expressions we find for the thermodynamic potential 
\eqn\Fvsugra{ F = - {V N^2 T^4 \over   (1-\Omega_1^2) (1-\Omega_2^2)} 
\left[\, {\pi^2\over 8} - {3\over 8 T^2}
\left(1-{\Omega_1^2+\Omega_2^2\over3}\right)  + O\left({1\over T^4}\right)\right] \, ,}
where $V=2 \pi^2$ is the volume of the unit $S^3$. We have used the 
AdS/CFT dictionary $G_5 = {G_{10}\over Vol(S^5)}={\pi\over 2 N^2}$.

\subsec{Summary of thermodynamics of four dimensional Kerr-AdS black holes}

The four dimensional Kerr-AdS metric has first appeared in \carter. 
It was subsequently studied in \hawkingi\ and \ckc. We take our conventions 
from \hawkingi\
 
\eqn\kerradsiv{ds^2 = -{\Delta\over \rho^2} \left( dt-{a\over \Xi} d\phi \right)^2 + {\rho^2
\over\Delta} dr^2 +  {\rho^2 \over\Delta_\theta} d\theta^2 +{\Delta_\theta\sin^2\theta \over \rho^2} \left( a dt - {(r^2+a^2)\over \Xi} \right)^2\,,}
with
\eqn\defsiv{\eqalign{
\Delta =& \; (r^2+a^2)(1+r^2)-2m r\, ,\cr
\Delta_\theta =& \; 1-a^2 \cos^2\theta\, ,\cr
\rho^2 =& \; r^2+a^2\cos^2\theta\, ,\cr
\Xi =& \; 1-a^2\,. }}
The case $m=0$ corresponds to empty AdS. 
Coordinates that exhibit the asymptotically AdS form are
\eqn\newcooriv{\eqalign{
t=t\, ,&\;\;y^2={r^2\Delta_\theta+a^2 \sin^2\theta\over\Xi}\,,\cr
y \cos\Theta =r \cos\theta\, ,&\;\;\Phi=\phi-a t\, .}}
The horizon radius $r_+$ is the largest root of $\Delta=0$.
The angular velocity relevant to the dual field theory thermodynamics is
\eqn\veloiv{\Omega = \Omega^H-\Omega^{\infty}={a (1+r_+^2)\over r_+^2+a^2}\,.}
The period of the Euclidean time coordinate is given by
\eqn\betaiv{ \beta = {1\over T} = {4\pi(r_+^2+a^2)\over \Delta^\prime(r_+)}\,.}
The Euclidean action calculated with respect to the AdS background is
\eqn\actioniv{ I =- {\beta (r_+^2+a^2)(r_+^2-1)\over 4 G_4 (1-a^2) r_+}\, ,}
which implies the existence of a phase transition at $r_+=1$.
Again we are interested in the high $T$ and large 
$r_+$ regime. The high
temperature expansions of $r_+$ and $\Omega$ are given by
\eqn\inverseiv{\eqalign{
r_+ = & \; {4\pi T\over 3} - {1-2\Omega^2\over 4 \pi T} + 
O\left({1\over T^2}\right) \, , \cr
a =& \; \Omega \left(1- {9 (1-\Omega^2)\over 16 \pi^2 T^2}\right) 
+O\left({1\over T^2}\right) \, .}}
Using this and the AdS/CFT dictionary for the M2-brane case,
$G_4 = {3\over 2\sqrt{2}} N^{-{3\over 2}}$, we find the following expression for the thermodynamic 
potential $F= T I$
\eqn\Fivsugra{F = - {8 \sqrt{2} \; V  N^{3\over 2} \,\pi^2\, T^3 \over 81 \, (1-\Omega^2)} \left[\, 1 - {9\,(2-\Omega^2)\over 16 \pi^2 T^2}+ O\left({1\over T^3}\right)
\,\right] \, .}
Here $V=4\pi$ is the volume of the unit two-sphere. 

\newsec{Field Theory Calculation}

We want to calculate the thermodynamics of four-dimensional
${\cal N}=4$ Yang-Mills on a rotating three-sphere in the
weak coupling limit. At weak (zero) coupling we will
calculate the 1-loop approximation to the partition function, in which
colored degrees of freedom run in the loop. This is only consistent
in the deconfined phase of the theory. The interacting ${\cal N}=4$
theory on the sphere is in a deconfined phase only at high temperature.
This holds even at weak coupling \refs{\witii, \bo}. Therefore we will
concentrate in the high temperature regime.

The thermal part of the 1-loop contribution to the partition function of 
a field theory on $S^3$ where matter is forced to rotate with angular 
velocities $\Omega_1$ and $\Omega_2$, is
\eqn\freeE
{\log Z = - \sum_i \; \epsilon_i \sum_{l,m_1, m_2}^{\infty} \log \Big( 1 - 
\epsilon_i \, e^{-\beta(\omega_l^i + m_1 \Omega_1 + m_2 \Omega_2)  } 
\Bigr) \; ,
} 
with $i$ labelling the particle species, $\epsilon_i=1$ for bosons and $-1$
for fermions, and $\omega_l^i$ is the energy of a mode of angular momentum 
$l$. We will assume that the scalars of ${\cal N}=4$ are conformally 
coupled since supergravity determines the field theory metric only up to 
conformal transformations. Then the thermodynamics will depend only on the 
dimensionless quantities $T R$ and $\Omega_i R$, where $R$ is the radius of 
the sphere. The high temperature
limit is equivalent to the large radius limit of the sphere keeping
constant the speed of the boundary, $\Omega_i R$. Thus the leading term
in the high temperature expansion of \freeE\ can be obtained
by substituting the discrete sums in \freeE\ by integrals \refs{\hawkingi, 
\berpar}. 
We will be interested in obtaining subleading terms; for that we will have 
to deal directly with the sums in \freeE.

\subsec{Approximate evaluation of sums}

As a first step we would like to evaluate sums of the generic form
\eqn\suma{
\sum_{l=l_1}^{l_2} f_l \; ,
}
where $f_l$ depends of an small parameter $\epsilon$ and we can define a 
function $f(x)$ such that $f_l= \epsilon f(\epsilon l)$. In the limit
of very small $\epsilon$, we have formally $\epsilon \Delta l \rightarrow dx$
($\Delta l=1$ in \suma). 
Our only hypothesis will be that $F(x)=\int \! f(x) \, dx$
is analytic in the interval $[x_1, x_2]$, with $x_1=\epsilon l_1$ and 
$x_2=\epsilon(l_2+1)$. We have then
\eqn\ecu{ 
\int_{x_1}^{x_2} \! f(x) \; dx = \sum_{l=l_1}^{l_2} \bigl[ 
F(\epsilon(l+1)) -F(\epsilon l) \bigr]
}
We can expand the rhs in a Taylor series,
obtaining
\eqn\ecutwo{
\sum_{l=l_1}^{l_2} \epsilon f(\epsilon l)= \int_{x_1}^{x_2} \! f(x)\; dx \; - 
\; \sum_{k=1}^{\infty} {\epsilon^{k+1} \over (k+1)!} \, 
\sum_{l=l_1}^{l_2} f^{(k)} (\epsilon l) \; ,
}
where $f^{(k)}=d^k f/ dx^k$. Applying recursively this reasoning
to $f^{(k)}$ we can evaluate \suma\ to any order in the small parameter
$\epsilon$. Up to order $\epsilon^4$ it is straightforward to obtain
\eqn\expan{
\eqalign{
\sum_{l=l_1}^{l_2} \epsilon f(\epsilon l)= & 
\left (F(x_2) - {\epsilon \over 2} f(x_2) + 
{\epsilon^2 \over 12} f'(x_2) - {\epsilon^4 \over 720} f'''(x_2)\right) \cr
& \;\;\;- \left (F(x_1) - {\epsilon \over 2} f(x_1) + 
{\epsilon^2 \over 12} f'(x_1) - {\epsilon^4 \over 720} f'''(x_1)\right) }
}
This expression is not the final answer since we will consider
situations in which both the functions that appear in \expan\ and 
$x_1,x_2$ can have an explicit dependence on $\epsilon$. In that case
we will have to further Taylor expand \expan\ to the desired order in
$\epsilon$.  

As a test of this approach, we will first 
calculate the high temperature expansion of the energy
of ${\cal N}=4$ Yang-Mills on $S^3$ without rotation at weak coupling.
This was obtained long ago in \old\ and recently reobtained using
the heat kernel method in \myers. The 1-loop contribution to the
thermal part of the energy of ${\cal N}=4$ Yang-Mills on $S^3$ is
\eqn\energyNR{
E= \sum_{s=0,1/2,1} \! \! n_s \;  \sum_{l=l_0^i}^{\infty} \; 
{d^s_l \omega_l^s \over 
e^{\beta \omega_l^s}-(-)^{2s}} \; ,
}
where $s$ denotes the spin of scalars, fermions and gauge bosons, $d^s_l$ is
degeneracy of particles with spin $s$ and  energy $\omega_l^s$ and 
$n_s$ are the number of fields of each spin present in the vector
multiplet of ${\cal N}=4$, i.e. $n_0=6$, $n_{1/2}=4$ and $n_1=1$. 
The data for scalars and fermions are
\eqn\data{
\eqalign{
s=0 \;: &\;\;  d_l= (l+1)^2 \;\;\; , \; l\geq 0 \; , \cr
s={1 \over 2} \; : & \;\;  d_l= 2 \Bigl( l+{1 \over 2} \Bigr) \Bigl( 
l+{3 \over 2}\Bigr) 
\;\;\; , \; l\geq {1 \over 2} \; . }
}
In both cases $\omega_l=(l+1)/R$ and $l$ runs on the integers for bosons and
half-integer for fermions. 

We have to take care of a subtlety in the spectrum of gauge fields
on the three-sphere at finite temperature. The spectrum
of a gauge theory on $S^3$ in the Feynman gauge
$\nabla^\mu A_\mu =0 $ has been found some time 
ago in \hoso. The vector field can be classified according to the
representation of the $SO(4)\cong SU(2)\times SU(2)$ isometry of the 
three-sphere. We label its representations by $(j_1,j_2)$.
The temporal
component is a scalar on the three-sphere and its modes fall 
into the $({l\over 2},{l\over 2})$, $l\ge0$  representation. The vector
modes on the sphere form
the $({l\over 2},{l\over 2}+1)$, $({l\over 2}+1,{l\over 2})$ and $({l\over
2}, {l\over 2})$, $l\ge 1$ representations. We denote the $({l\over
2},{l\over 2}+1)$  and $({l\over 2}+1,{l\over 2})$
fields by $\vec{A}_\pm$ and the $({l\over 2}, {l\over 2})$ modes of the
three-vector by $\vec{A}_0$.
The energies are given by
\eqn\vecmodes{\eqalign{
A_0 \; :& \; \; \omega_l = {\sqrt{l(l+2)}\over R}\;\;\; , \; l\ge 0\;,\cr
\vec{A}_0 \; :&\;\;  \omega_l = {\sqrt{l(l+2)}\over R}\;\;\;, \; l\ge 1\;,\cr
\vec{A}_\pm \; :&\;\; \omega_l = {l+1\over R} \;\;\; , \;l\ge 1\;.}}
In addition we have the modes coming from the ghosts. They are (minimally
coupled) scalars and therefore fall into the  $({l\over 2},{l\over 2})$
representations. 
\eqn\ghostmodes{\eqalign{ c, \bar{c}\; : &\;\; \omega_l = {\sqrt{l(l+2)}\over
R}\;\;\; , \;l\ge 0\;.}} 
The ghost fields have to be taken periodic around the $S^1$ and thus are
subject to the Bose-Einstein distribution despite their fermionic nature.
They contribute with negative sign to \energyNR.
We see now that the ghosts compensate the $A_0$ and the $\vec{A}_0$
contributions up to a left over zero energy mode.
If we were on $R\times S^3$ we could gauge away the zero mode of 
the temporal component of the gauge field and there would be no reason
to
include the zero modes of the ghosts. On $S^1\times S^3$ however
the gauge field zero mode can not be removed by a proper (periodic on
$S^1$)
gauge transformation. Therefore these zero modes have to be included in
the
spectrum. The contribution from the left over zero mode to \energyNR\ is
\eqn\zeromode{ E = - \lim_{\omega \rightarrow 0} \; {\omega \over e^{\beta\omega}-1} =
-{1\over\beta}\,.}   
We have to add this to the contributions from the transversal gauge field
modes.

After these remarks we can apply \expan\ to \energyNR.
The small parameter is $\epsilon=1/TR$. 
The function $f_i(x)$ associated to each species can be obtained simply 
by substituting $l \rightarrow x/\epsilon$ in \energyNR. The upper limit 
of the continuous 
variable is $x_2=\infty$. All functions appearing in the rhs of \expan\ 
tend to zero at infinity \foot{The function $F$ is only defined up to a
constant. This constant can be taken such that $F(x\rightarrow
\infty)=0$.}, 
therefore the sum is determined by the
value of the functions at $x_1$. We have $x_1=0,\epsilon/2,\epsilon$ for 
scalars, fermions and (the transversal modes of the) gauge bosons
respectively, obtaining
\eqn\expanNR{
E= - T^4 R^3 \left[ \sum_{s=0,1/2,1} n_s \, \left( F_s (\epsilon s) - 
{\epsilon \over 2} f_s (\epsilon s) + 
{\epsilon^2 \over 12} f'_s(\epsilon s) - {\epsilon^4 \over 720} 
f'''_s (\epsilon s) \right) + O(\epsilon^5) \right],
}
where all functions are well defined in the limit $\epsilon \rightarrow
0$.
We have now to Taylor expand each function around $\epsilon=0$. The 
definition \freeE\ involves logarithms which make 
Taylor expansions around $\epsilon=0$ ill-defined. This is the reason why
we 
chose to calculate the energy instead of $\log Z$. We can then use the
relation $E=-{\partial \over \partial \beta} \log Z$ to derive
the partition function up to a temperature independent term.
From \expanNR\ we obtain the contribution of each species to the energy 
of the gauge theory \refs{\old, \myers}
\eqn\eneNR{
\eqalign{
s=0\;: &  \; 6 \; V \left( \, { \pi^2 T^4 \over 30}\, - \,{1 \over 480
\pi^2 R^4} 
\, \right) \; , \cr
s={1 \over 2} \;: & \;  4 \; V \left(\, {7 \pi^2 T^4 \over 120} \, - \,
{T^2 \over 48 R^2} \, - \,
{17 \over 1920 \pi^2 R^4} \, \right) \; , \cr 
s=1\;: & \;\;\;\;  V \left( \,{\pi^2 T^4 \over 15} \, - \, {T^2 \over 6
R^2} \, + \, 
{T \over 2 \pi^2 R^3} \, - \, { 11 \over 240 \pi^2 R^4} \, \right) \; , }
}
where $V=2 \pi^2 R^3$ is the volume of $S^3$. 
The term at order $\sim T/R^3$ for the gauge bosons is precisely
cancelled by the contribution of the left over ghost zero mode \zeromode.
The terms independent of the temperature equal minus the 
Casimir energy of each field on the sphere. The final result for the
energy 
is 
\eqn\resultNR{
E= V \left( {\pi^2 T^4 \over 2} - {T^2 \over 4 R^2} \right) - \, E_{C}\, +
\, O\Bigl( {1 \over T R^2} \Bigr) \; ,
}
with $E_C$ denoting the Casimir energy of ${\cal N}=4$ Yang-Mills
on a three-sphere.

\subsec{${\cal N}=4$ Yang-Mills on a Rotating 3-sphere}

After having tested our method on a simple example, we want to apply it
to ${\cal N}=4$ Yang-Mills on a rotating three-sphere. The partition function 
can be used to obtain 
the thermodynamic potential $F=-T \log Z$ in the grand-canonical ensemble,
where the thermodynamic variables are the temperature and the angular 
velocities. As before, we will calculate ${\partial \over \partial \beta} 
\log Z$ instead 
of $\log Z$. We set in the following $R=1$ for convenience. 
Therefore the small parameter will be just the inverse of the 
temperature, $\beta$. We will analyse separately the 
contribution of scalars\foot{Expression (3.11) gives twice the contribution of 
a real scalar. Notice that for $s=0$ the summation is already symmetric under 
the interchange of $\Omega_+$ and $\Omega_-$.}, fermions and gauge bosons, which in 
terms of the spin of each species is given by \hawkingii\
\eqn\calR{
-{ \partial \log Z \over \partial \beta} \; = \; \sum_{l=s}^{\infty} \; \sum_{m=-{l+s \over 2}}^{{l+s \over 2}} \;  \sum_{n=
-{l-s \over 2}}^{{l-s \over 2}} \; {l+1 + m \Omega_+  + n \Omega_- \over
e^{\beta(l+1 + m \Omega_+  + n \Omega_- )} - (-)^{2 s} } \;\; + \;\; 
( \Omega_+ \leftrightarrow \Omega_-) \; .
}
Here $\Omega_\pm$ are the angular velocities corresponding to the Cartan
elements of the $SU(2)\times SU(2)$ rotation group of the
three-sphere.
The angular velocities \velosv\ are related to them by
$\Omega_\pm=\Omega_1\pm\Omega_2$ \hawkingii.

We will only be interested in the leading and first subleading term
of the high temperature expansion. For that it is enough to evaluate
the sums to order $\beta^2$. We have to apply three times \expan. 
The calculation simplifies by the fact that we can define a single
variable $x$ such that when $x=\beta(l+1)+\beta m\Omega_+ +\beta n\Omega_-$
the function 
\eqn\funct{
f(x)= {x \over e^x - (-)^{2s}} \; , 
}
equals $\beta f_{lmn}$, where $f_{lmn}$ denote the
summand in \calR.
The function $f(x)$ and its integrals are analytic between $[0,\infty]$. 
The result of the $n$ summation is 
\eqn\nsum{
\sum_n f_{lmn}= T^2 \sum_{i=1,2} (-)^i \left[ {1 \over \Omega_+} F(x_i) -
{\beta \over 2} f(x_i) + {\beta^2 \over 12} \Omega_+ f'(x_i) \right] \; ,
}
where $x_2= \bigl( \beta(l+1)+\beta m\Omega_+ +\beta({l-s \over 2}+1)
\Omega_- \bigr)$, $x_1=\bigl( \beta(l+1)+\beta m\Omega_+ +
\beta(-{l-s \over 2})\Omega_- \bigr)$, 
$F=\int \! f dx$ and $f'=df/dx$. Defining integrals and derivatives with
respect to $x$ instead of the continuous variable associated to $n$ is 
the reason for the extra factors of $\Omega_+$. 
It is equally straightforward to perform the summation in $m$, obtaining 
\eqn\nmsum{
\eqalign{
\sum_m \sum_n f_{lmn} = & \; \;
T^3 \sum_{i,j=1,2} {(-)^{i+j} \over \Omega_+ \Omega_-}
 \biggl[ \, h(x_{ij}) 
- {\beta \over 2} \, ( \Omega_+ + \Omega_-) \, F(x_{ij}) \cr  + & 
{\beta^2 \over 12} \, (\Omega_+^2+3\Omega_+\Omega_- +\Omega_-^2) \, f(x_{ij})
\, \biggr]
\; ,}
}
where $h=\int \! F dx$ and the four values $x_{ij}$ coming from the upper and 
lower bound of the two summations are
\eqn\zs{
\eqalign{
x_{ij}= &\;\; \beta \, \alpha_{ij} \, l + \beta \, \gamma_{ij} \; , 
\;\;\;\;\;\;\;\;\;\;\;\;\;\;\;\;\;\;\;\;\;\;\;\;\;\;\;\;\;\;\;\; 
i,j=1,2 \; , \cr
\alpha_{ij}= & \;\; 1 + {(-)^i \Omega_+ + (-)^j \Omega_- \over 2} \; , \cr
\gamma_{ij}= & \;\; 1 + {1+ (-)^i (1+s) \over 2} \, \Omega_+ +
{1 + (-)^j (1-s) \over 2} \, \Omega_- \; .}
}
Now we can perform the last summation. The functions $H=\int \! h dx$, $h$ and 
$F$ are defined only up to constants. We will choose these constant such that
all of these functions vanish as $x \rightarrow \infty$. The initial function $f$ 
\funct\ also vanishes at infinity. Therefore the triple sum is 
\eqn\lnmsum{
\eqalign{
\sum_l \sum_m  \sum_n & \; f_{lmn} \; = \; T^4  
\sum_{i,j=1,2} {(-)^{i+j} \over \Omega_+ \Omega_-}
 \Biggl[-{1 \over \alpha_{ij} } H( \beta \delta_{ij}) 
+ {\beta \over 2} \biggl( 1 + {\Omega_+ + \Omega_- \over \alpha_{ij}}\biggr) 
h(\beta \delta_{ij})  \cr
& \;\;\; - 
{\beta^2 \over 12} \biggl(\alpha_{ij} + 3 (\Omega_+ + \Omega_-) +
{\Omega_+^2+3\Omega_+\Omega_- +\Omega_-^2 \over \alpha_{ij}} \biggr) 
F( \beta \delta_{ij}) \Biggr]  \; ,} 
}
with $\delta_{ij}= \alpha_{ij} s + \gamma_{ij}$. The last step consists
in expanding each function around $\beta=0$ and finally adding the same 
expression with $\Omega_+$ and $\Omega_-$ interchanged. The coefficient 
of the leading $T^4$ term is simply given by
\eqn\leading{
- {2 H(0) \over \Omega_+ \Omega_-} \sum_{i,j=1,2} {(-)^{i+j} \over \alpha_{ij}}
\; .
}
The factor $2$ appears because \leading\ is symmetric under the interchange
of $\Omega_+$ and $\Omega_-$. The next order contribution is
\eqn\ordera{
{h(0) \over 2 \, \Omega_+ \Omega_-} \sum_{i,j=1,2} (-)^{i+j}
\left[ \, { -2 \delta_{ij} +\Omega_+ +\Omega_- \over \alpha_{ij}} \; +
\; ( \Omega_+ \leftrightarrow \Omega_-) \, \right]  \; .
}
Using \zs\ one can see that \ordera\ is equal to zero. At the order $T^2$ 
we obtain the following expression
\eqn\orderaa{
\eqalign{
- {F(0) \over 12 \Omega_+ \Omega_-}  \sum_{i,j=1,2} (-)^{i+j} &
\biggl[\, {6 \, \delta_{ij}^2 \over \alpha_{ij}} \, - \, 6 \, \delta_{ij} 
\Bigl(1 + {\Omega_+ + \Omega_- \over \alpha_{ij}}\Bigr) \cr
&\;\;\;\; + \alpha_{ij} + {\Omega_+^2 +3 \Omega_+ \Omega_- + \Omega_-^2 
\over \alpha_{ij}}  \; + \; ( \Omega_+ \leftrightarrow \Omega_-) \, \biggr]  \; .}
}
Integrating \calR\ we obtain the contribution to the thermodynamical 
potential of real scalars, Weyl fermions and gauge bosons living 
on a rotating 3-sphere
\eqn\FR{
\eqalign{
F_{s=0}=\;& {V \over (1-\Omega_1^2)(1-\Omega_2^2)} \, \left[ \,-{\pi^2 T^4 \over 90} \, + \, 
{T^2 \over 72}(\Omega_1^2 + \Omega_2^2)\, \right]\; , \cr
F_{s=1/2}=\; & {V \over (1-\Omega_1^2)(1-\Omega_2^2)} \, \left[\,- {7 \pi^2 T^4 \over 360}\, + \, 
{T^2 \over 48} \Bigl(1- {\Omega_1^2 + \Omega_1^2 \over 3}\Bigr) \,  \right]\; , \cr
F_{s=1}= \;  & {V \over (1-\Omega_1^2)(1-\Omega_2^2)} \, \left[ \,- {\pi^2 T^4 \over 45} \, + \,
{T^2 \over 6} \Bigl(1- {5(\Omega_1^2 + \Omega_2^2) \over 6}\Bigr) \,  \right]\; . }
} 
We have used $\Omega_{\pm}=\Omega_1 \pm \Omega_2$. For the particular case of
${\cal N}=4$ Yang-Mills we have
\eqn\freeRYM{
F_{{\cal N}=4}= {V N^2 \over (1-\Omega_1^2)(1-\Omega_2^2)} \, \left[ \,-{\pi^2 T^4 \over 6} \, + \, 
{T^2 \over 4}\Bigl(1-{\Omega_1^2 + \Omega_2^2 \over 3}\Bigr)\, \right]\; .
}
In \berpar\ the leading contribution to the thermodynamics of 
${\cal N}=4$ Yang-Mills on a rotating $S^3$ was analysed in the high 
temperature limit. The discrepancy by a 
factor $3/4$ between the strong and weak coupling limits of the 
thermodynamics potentials which holds for the gauge theory in flat 
non-rotating space, was found in \berpar\ to hold exactly also for the case 
with one rotation parameter, i.e. $\Omega_2=0$. This is a somehow surprising
result since the rotation introduce new dimensionless parameters such 
that strong and weak coupling limits could differ by an arbitrary
function of the $\Omega$'s. Comparing \Fvsugra\ and \freeRYM\ we observe
that the simple factor $3/4$ between the $T^4$ term at strong and weak 
coupling persists in the generic case with two rotations. 
Moreover, a stronger result holds. The strong and weak coupling 
contribution to the thermodynamic potentials at the subleading order 
$T^2$ differ 
again only by a numerical factor, $3/2$. There is no obvious symmetry
reason that could explain this behaviour.  

The relation between strong and weak coupling limits of the thermodynamic
potential 
for the system with two rotations as a function of the horizon radius 
$r_+$ was investigated numerically in \hawkingii. The ratio $F_{st}/F_w$ 
decreases from $3/4$ at $T \rightarrow \infty$ to zero at the Hawking-Page 
phase transition point $r_+=1$. However it decreases slower for systems 
with stronger rotation. We can reproduce that result for high temperature 
from \freeRYM\
\eqn\quo{
{F_{st} \over F_w} = {3 \over 4} \left[ \, 1 - {3 \, \beta^2 \over 2\pi^2 }\Bigl(
1 - {\Omega_1^2 + \Omega_2^2 \over 3} \Bigr) \, \, \right] \; .
} 
The term in parenthesis varies between $1$ in the non-rotating case
and $1/3$ when both $\Omega_1,\Omega_2 \rightarrow 1$.

\subsec{The \neight\ supersingleton on a rotating sphere}

As already mentioned in the introduction, the four dimensional Kerr-AdS black hole is thought
to be dual to the superconformal field theory arising on coinciding M2-branes. 
We only have a Lagrangian formulation for the case of the field theory on a single M2-brane.
This is the \neight\ supersingleton theory consisting of eight conformally coupled scalars and
eight spinors. It seems most natural to compare the large $N$ limit of this theory represented
by supergravity to the $N=1$ field theory, since there is no dimensionless 
coupling as in 
the four dimensional case. For the same reason as before we will actually compute
${\partial\over \partial \beta} \log Z$ 
\eqn\calRiii{ - {\partial\over \partial \beta}\log Z =
\sum_{l=s}^\infty\,\sum_{m=-l}^l\, {l+{1\over 2}+m\Omega\over e^{\beta(l+
{1\over 2}+m\Omega)}-(-)^{2s}}\,}

The calculation parallels the previous case. Therefore we do not present the
details.
We obtain for the thermodynamic potential
\eqn\FreeRSingleton{F \; = \; -{V \, T^3 \over 1-\Omega^2}  
\left[\, {7 \zeta(3)\over \pi} +
{1-2\Omega^2\over 6\pi} {\log(T)\over T^2} \, \right]\,.}
It has been noted in \kt\ that the leading term contains a $\zeta$-function with odd integer argument.
For this reason there can not be an agreement up to some rational number between the
supergravity result and the field theory result. For the leading term 
there is still a minimal agreement between the two results since their scaling 
with the temperature is the same.
This is however to be expected because the leading term corresponds to the flat space limit and in
this case conformal symmetry dictates the scaling with $T$. The scaling of the subleading term can
however not be predicted by conformal symmetry. As we see now it is indeed different. The field 
theory calculation shows logarithmic behaviour, whereas the supergravity result gives polynomial
behaviour\foot{We remind the reader that we have set the radius of the sphere
$R=1$. In this conventions $T$ coincides with the dimensionless parameter $\epsilon$ introduced earlier.}. In fact even the functional dependence on $\Omega$ is different for the subleading
terms. In the light of these results it seems even more remarkable that the four dimensional
calculations show such a high agreement.

\vskip1cm

\centerline{\bf Acknowledgments}
\vskip2mm
We would like to thank E. Alvarez and M. Tytgat for discussions.

\listrefs
\end